# Somatic Practices for Understanding Real, Imagined, and Virtual Realities

**Lisa May Thomas**
Intangible Realities Laboratory
University of Bristol, Bristol, UK
lt15256@bristol.ac.uk

**Helen M. Deeks**
Intangible Realities Laboratory
University of Bristol, Bristol, UK
h.deeks@bristol.ac.uk

**Alex J. Jones**
Intangible Realities Laboratory
University of Bristol, Bristol, UK
alex.j.jones@bristol.ac.uk

**Oussama Metatla**
Department of Computer Science
University of Bristol, Bristol, UK
o.metatla@bristol.ac.uk

**David R. Glowacki***
Intangible Realities Laboratory
University of Bristol, Bristol, UK
glowacki@bristol.ac.uk

## ABSTRACT

In most VR experiences, the visual sense dominates other modes of sensory input, encouraging non-visual senses to respond as if the visual were real. The simulated visual world thus becomes a sort of felt actuality, where the 'actual' physical body and environment can 'drop away', opening up possibilities for designing entirely new kinds of experience. Most VR experiences place visual sensory input (of the simulated environment) in the perceptual foreground, and the physical body in the background. In what follows, we discuss methods for resolving the apparent tension which arises from VR's prioritization of visual perception. We specifically aim to understand how somatic techniques encouraging participants to 'attend to their attention' enable them to access more subtle aspects of sensory phenomena in a VR experience, bound neither by rigid definitions of vision-based virtuality nor body-based corporeality. During a series of workshops, we implemented experimental somatic-dance practices to better understand perceptual and imaginative subtleties that arise for participants whilst they are embedded in a multi-person VR framework [1]. Our preliminary observations suggest that somatic methods can be

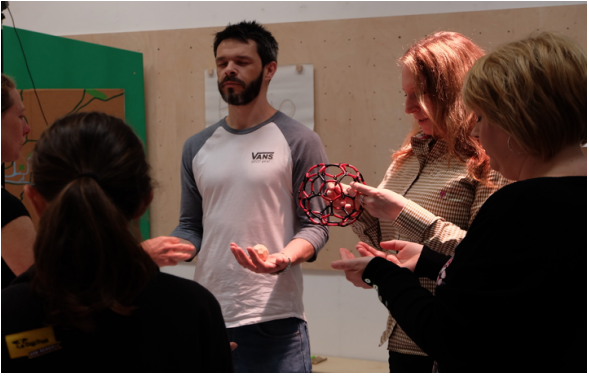

Participants with their eyes closed try and understand abstract structures using only somatosensory mechanisms, without recourse to vision used to design VR experiences which enable (i) a tactile quality or *felt* sense of phenomena in the virtual environment (VE), (ii) lingering impacts on participant imagination even after the VR headset is taken off, and (iii) an expansion of imaginative potential.

## 1 THE VR 'PERCEPTION GAP'

There is a sense in which the VR headset or head-mounted display (HMD) acts like a blindfold, blocking out all visual information of the 'world out there'. But it is more complex than simply having your eyes closed or blindfolded, because the HMD enables a more dominant visual virtual experience to 'take over' the sensory domain. Perceptually, the visual VE is layered onto the physical environment. In many cases, the embodied felt sense within the physical environment does not necessarily correlate with what the simulated VE is encouraging you to think. This 'perception gap' can lead to a disruption between visual and felt senses, but it also mediates continuous dialogue between the real (body) and the virtual (environment) in an effort to navigate the gap [2]. For example, perceptual mechanisms which inform bodily responses to the VE can override one's cognitive understanding that the simulated virtual world is not real [3]. This is linked to the fact that vision overrides other sensory information. The perceptual gap arises from the fact that what is seen as *virtual* layers over what is felt as *actual* or *physical* [2]. However, in moments where the visual virtual world is disturbed, one's relationship with the physical environment re-emerges. It is in these moments that the gap in perception is highlighted: for example, in moments when one reaches out to touch something that can be seen in the VE but which isn't physically there, or when one touches something which is physically present but which cannot be seen in the VE.

## 2 BRIDGING THE GAP: HAPTICS AND VR

In order to better integrate layers of perception, and overcome the aforementioned 'perceptual gap', many researchers are developing tools and tactics in order to allow users in VR to not only see and hear, but also to feel virtual worlds, aligning sensations like tactile and touch sensations with the simulated VE [4]. For example, Michael Abrash, the chief scientist at Oculus VR, has been cited as saying: "As important as haptics potentially is for VR, it's embryonic right now. There's simply no existing technology or research that has the potential to produce haptic experiences on a par with the real world" [5]. Certainly, haptics has not met (in the same way that visual display technologies have) the "blueprint" laid out in "The Ultimate Display" by Sutherland in 1965 [6]. Whereas vision and sound can be used to produce immersive effects, "Feeding the complex and variegated data of touch back… [has] proved nearly impossible", as "The haptic system [has] resisted translation into machine-legible code" [5]. Many technologies developed to encourage a tactile sense rely on the visual dominance and its corresponding expectations for what things which are seen might feel like. For example, consider a case where a user is holding a tool in VR (e.g., a light-saber). If they use the light-sabre to say, deflect a laser beam, they will anticipate a tactile sensation (e.g., a vibration), which can be provided via vibrators mounted in the VR controllers. In this case, the controller's vibration, which has limited tactile possibilities, derives meaning only alongside the visual input. To circumvent this difficulty, some approaches embed VR within controlled physical environments, so that actual physical objects (which offer fuller tactile experiences) map to their simulated virtual (visual) counterparts.

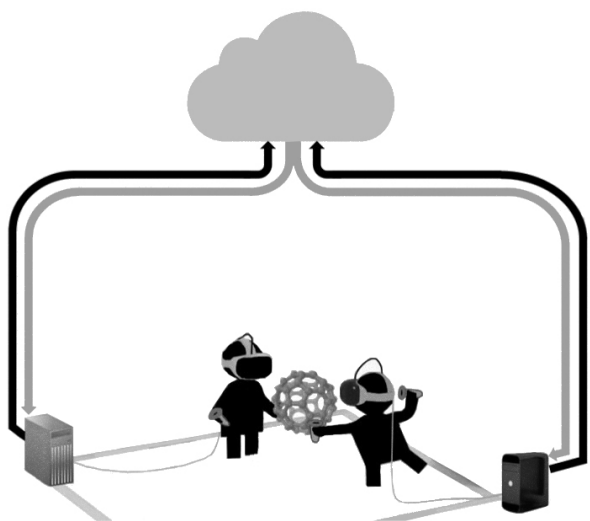

Figure 1: schematic of the multi-person VR framework utilized in this work, showing two users interacting with a real-time dynamical simulation of a $C_{60}$ molecule, which is hosted on a computational physics server.

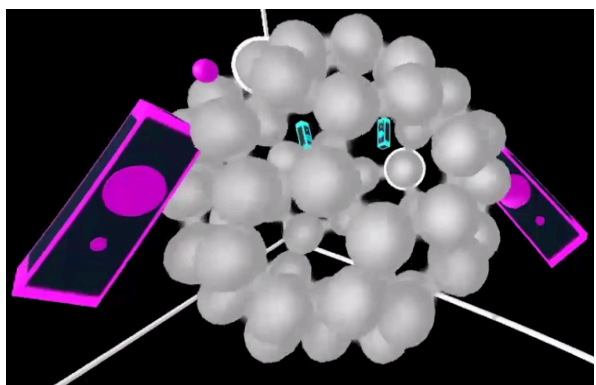

Figure 2: first-person view from one participant's perspective as they use their rose-colored controllers to pass a dynamical $C_{60}$ structure to another participant, whose controllers are shown in blue. A more detailed video is shown at https://vimeo.com/244670465

## 3 SOMATICS IN HCI

Somatic practices are widely present and growing in popularity across contemporary society, supported by a growing body of research demonstrating how they can be applied to support both physical and psychological health. For example, somatic practices which incorporate mindfulness, movement forms, visualizations, and breath exercises, are designed to enable an individual to 'check in' and observe their interior landscape (e.g., their awareness, their state of mind, their breath, and their body). Sensory practices like these are often designed to allow practitioners to perceive subtle relationships between movement, breath, gravity, body position, weight distribution, bodily tension, etc. Somatics, a term coined by Thomas Hanna, is defined as "the field which studies the soma: namely the body as perceived from within by first-person perception" [7]. Somatic practices are used frequently in dance training as methods which de-emphasises visual form to encourage expanded ways of seeing, re-adapting the sensory system toward modes of perceiving one's own body, other bodies, and space, thereby freeing up visually imposed boundaries and generating the capacity to experience felt body-to-body and body-to-environment connections [2]. Amanda Williamson writes about "The advent and growth of somatic movement/dance modalities, with their unique co-extensive emphasis on the deep *felt* interiority and connective external life (e.g., socio-communal, ecological, and political aspects) of the body" [8]. Recent work in HCI has linked the growing participation in contemporary somatic practices with current and emerging trends in HCI communities, specifically as a means for informing interaction design: "More and more, interaction design is incorporating values of self-observation, autoethnography, somaesthetics, and first-person methodologies in the design of personal, public, social, and everyday technologies" [9]. For the vast majority of the HCI field, attempts to improve interactive experiences are generally focussed on developing and improving the technology *per se*. With a more detailed understanding of the somatic mechanisms whereby existing HCI technologies achieve their effects on our human perceptual systems, it may be possible to imagine a scenario where HCI instead focuses its efforts on honing and training human perceptual sensitivities.

## 4 INVESTIGATING VR USING SOMATIC PRACTICES

The aim of this research is to utilize somatic practices as a research lens for better understanding whether tactile and haptic sensitivities operate in VR experiences *without* the need to employ sophisticated (and often costly) haptic devices to simulate (or stimulate) a sense of touch. Using somatic practices enables us to gain insight into those subtle perceptual sensitivities which operate when a person is simultaneously embedded in a virtual and physical environment. The open-source VR framework used herein is one which we have been developing over the past few years. Utilizing HTC Vive clients which connect to a real-time molecular physics server, it enables multiple users to move and interact with the structures and dynamics of complex molecular structures 'on the fly' and to interact with other users in the same VE [1], as shown in Figures 1 & 2. To date, this open-source framework (https://gitlab.com/intangiblerealities) found significant application in scientific research & education. Over time, we have noted consistent reports from users remarking how they are able to 'feel' differences between different molecular physics simulations, despite the fact that we have not implemented any haptic feedback in our system. In an attempt to better understand the perceptual mechanisms at work when users refer to different molecular physics simulations 'feeling' differently, we have used dance-somatic practices as a design tool for workshops where groups of participants journey through several

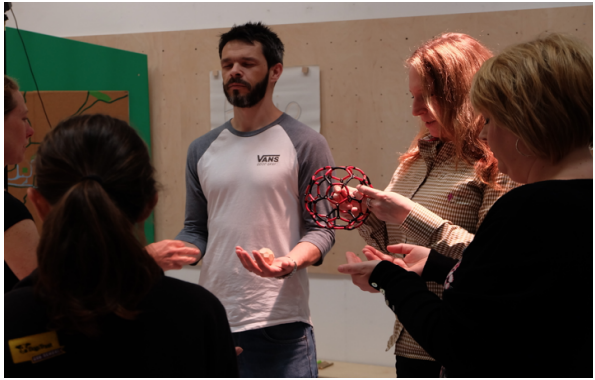

Figure 3: Sensory stage S1 of the workshop, where users (here shown with their eyes closed) were given tangible physical structures.

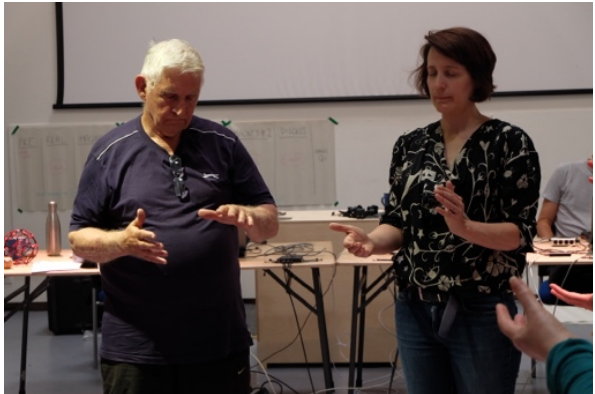

Figure 4: Sensory stage S2 of the workshop, where users (again shown with their eyes closed) were instructed to imagine ball-shaped dynamical structures.

'stages' together. In designing these workshops, we use the unique capacity of our VR framework to enable (i) interaction and manipulation of 3D virtual, dynamic forms, which have little analogue to anything which a participant would experience in the 'actual' world; and (ii) shared VR experiences, in which participants are virtually and physically co-present together.

## 5 EXPERIMENTAL WORKSHOPS COMBINING VR WITH SOMATICS

### 5.1 Workshop Design

A series of experimental workshops were designed and offered to public participants in April and May 2018 at Knowle West Media Centre in Bristol. As shown in Figures 3 – 7, groups of participants journeyed through six sensing stages to explore those sensations which arose from interacting with buckminsterfullerene ($C_{60}$), a ball-shaped molecular structure (whose discovery was awarded the 1996 Chemistry Nobel Prize). During the workshops, participants were guided through the following sensory stages, each of which encouraged users to engage with a different $C_{60}$ representation:

- Stage (S1) Participants were given tangible physical structures (Figure 3)
- Stage (S2) Participants were instructed to imagine 'ball-shaped' dynamical structures (Figure 4)
- Stage (S3) Participants were invited to interact with a real-time $C_{60}$ simulation whose dynamics were sonified. Users were not in VR, so that there was no corresponding visual component (Figure 5)
- Stage (S4) Participants were invited to experienced real-time interactive VR simulations of dynamical $C_{60}$ structures without the corresponding sonification (Figure 6)
- Stage (S5) Participants were invited to experienced real-time interactive VR simulations of dynamical $C_{60}$ structures with the corresponding sonification (Figure 6)
- Stage (S6) Participants were invited to re-imagine the dynamical $C_{60}$ structures (Figure 7)

The four sensing stages (S1), (S3), (S4), and (S5) utilized identical 'ball and stick' $C_{60}$ structural representations, although in each case these were experienced using dramatically different senses. For parts (S2) and (S6), users were asked to 'imagine a ball', given the ball shape of $C_{60}$. During each of the four sensing stages, participants were invited to "notice how the objects feel" and "how you sense them". All of stages except for the VR stages (S4) and (S5) were done with and without sight. After each stage, there were a few minutes for participants to write or draw their immediate thoughts. At the end of the workshop, they were invited to respond to a series of questions in the group. A significant aspect of the study design involved deciding stage ordering. We sought to provide an initial physical framework for the object in (S1), and then draw on participants' imaginations in (S2). After this we wanted participants to explore the virtual structure, offering the sonified VR experience (S3) prior to the visual VR experience in (S4) and (S5), as we know that vision dominates the sensory system. Finally, we were interested to understand the extent to which this experience of VR had resided or not in the imaginations of the participants (S6).

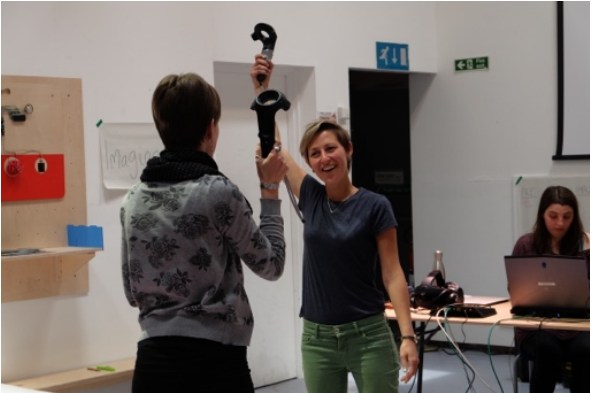

Figure 5: Sensory stage S3 of the workshop, where users (here shown with their eyes open) were instructed to interact with the real-time $C_{60}$ dynamics simulation shown in Fig 1, but using only audio feedback.

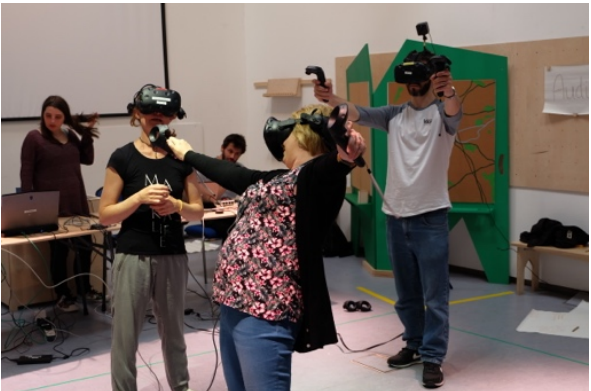

Figure 6: Sensory stages S4 and S5 of the workshop, where users were led into the VR system (shown in Fig 1) and instructed to interact with the real-time $C_{60}$ dynamics simulation, using different combinations of audio and visual feedback described in the text.

### 5.2 Workshop Themes

There were three high-level themes which emerged from the participant feedback:
- Theme 1 concerns how participants connected to each other across the stages, and how they connected to the $C_{60}$ structures offered as both physical and virtual - i.e. as physical $C_{60}$ objects which were handed to them and as simulated molecular structures in the VE.
- Theme 2 is about human and technological barriers which caused issues for participants as they attempted to 'map' processes between different modalities, for example moving between: (i) one sensory mode to another, e.g., working with eyes closed and then open; and (ii) one model of reality to another, e.g., interacting with an actual and then virtual $C_{60}$.
- Theme 3 concerns imagined realities, highlighting participants' perceptions of the bounds of their own imaginations, both in the physical space of the workshop and in the VE. This is a larger and more complex theme which was composed of two sub-themes: (i) The role that social and cultural norms play in bounding, enabling, and disabling imagination; and (ii) the shift made by the participants in terms of their perceived sense of their imaginative capabilities as they moved from the initial imagined stage (S2) to the latter stage (S6).

### 5.3 Discussion of Themes

In all three themes, there is the sense of a shift from one perceived state or mental model to another. In the first theme, participants initially struggled to connect to other participants and to the $C_{60}$ structures, but later in the workshop their sense of self in relation to each other and in relation to the $C_{60}$ structures shifted. Specifically, they shifted from an awkwardness (in the first couple of stages) to a sense of becoming more porous beings through whom virtual structures could move (in the VE), until they were finally able to imagine and move with malleable imagined forms in the final stage (S6). Participants also commented on the fact that the multi-person VR environment – particularly when it was it was not characterized by a specific object, e.g., in stage (S4) – operated as a social space with the ability to mediate new and different social conventions amongst groups of participants, and between a person and non-human 'other'. In the second theme, we identified the barriers which impacted participants' experiences. These barriers – both technological and human – appeared to hinder a range of mapping processes. For example, we noticed a barrier in moving between actual and virtual: specifically, many participants remarked that the HTC Vive controllers operated as a barrier to the transference of a 'felt' or tactile sense of the interactive dynamical structures simulated in the VE. We also noticed a barrier in moving between visual and tactile: specifically, participants remarked that the use of somatic practices enabled them to attend and listen to their own moving, sensing bodies, and allowed time for transitions between modes of perception to occur in ways which were personalized to each participant. For example, somatic encouragement during stages (S1) and (S2) led to some participants using these tools during subsequent stages where attention to subtlety was required – e.g., during stage (S3). In the third theme, the issue of a 'bounded-imagination' arose – highlighting limitations in the participants' capacities to imagine and also limitations of the imagination itself, including its social and cultural boundedness and its bias toward things which can be seen and are regularly encountered in day-to-day experience. The last workshop stage (S6) enabled participants to open up their imagination, extending its possibilities, and transcending the borders of the familiar. For example, some participants during stage (S2) stated that they 'didn't have an imagination' or that 'imagination is for kids'. However, when they were asked to re-imagine in stage (S6),

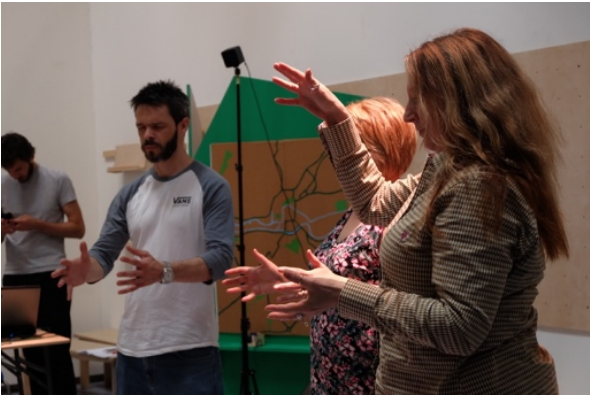

Figure 7: Sensory stage S6 of the workshop, where users (again shown with their eyes closed) were instructed to re-imagine ball-shaped dynamical structures.

## ACKNOWLEDGMENTS

We thank the Knowle West Media Center (Bristol, UK) for providing us space to carry out these public workshops. LMT acknowledges funding from Arts Council England, the Royal Society, and the Leverhulme Trust. H.M.D. and A. J. J. are funded through PhD studentships from the EPSRC. O.M. acknowledges funding as an EPSRC fellow from grant EP/N00616X/2. DRG acknowledges funding as a Royal Society Research fellow and Philip Leverhulme fellow, with additional support from EPSRC programme grant EP/P021123/1, and the London Barbican.

the somatic practices combined with the VR experience enabled them to be comfortable in using their imaginations. For others, it was clear to us from the feedback that the somatic and VR journey encouraged them to move from more familiar everyday imaginings which where separate from their bodies (e.g., my son's red ball) toward more abstracted, malleable, energetic, and porous forms which they were able to conceive being continuous with their body.

## 6 CONCLUSIONS

The use of somatic practices as tools for analyzing and designing VR experiences can support participants in attending to subtle shifts and changes in their perception of self and other, particularly as they moved between significantly different sensory modes. These somatic processes take time, but allowed participants to challenge normative modes for how they understand phenomena and the connection between things. Somatic practices also provide important insights into the subtleties of VR experiences, for example enabling us to gather participant evidence that the controllers operate as a potential barrier to 'feeling' the dynamics of the tactile structures which were being simulated in the VE. The sense of being in the VE, and the presence of the forms within the VE, clearly had residual effects after participants removed their headsets, impacting their subsequent ability to imagine, and enabling people to go beyond the borders of those culturally-bounded norms and conventions which structure their imagination.

## REFERENCES

[1] M. O'Connor *et al*., 2018, Sampling molecular conformations and dynamics in a multiuser virtual reality framework, *Science Advances,* vol 4, no 6, eaat2731, http://doi.org/10.1126/sciadv.aat2731
[2] M. Slater, 2010, First Person Experience of Body Transfer, *Plos One*, http://doi.org/10.1371/journal.pone.0010564
[3] L. M. Thomas and D. R. Glowacki, 2018, Seeing and feeling in VR: bodily perception in the gaps between layered realities, *Int. J. Performance Art and Digital Media*, vol 14, no 2, http://doi.org/10.1080/14794713.2018.1499387
[4] L. P. Chen, S. Marwecki, P. Baudisch. 2017. Mutual Human Actuation, In *Proceedings of the 30th annual ACM symposium on User interface software and technology (UIST '17).* http://doi.org/10.1145/3126594.3126667
[5] D. Parisi, 2018 Archaeologies of touch: Interfacing with Haptics from Electricity to Computing. http://doi.org/10.5749/j.ctt20mvgvz
[6] I. E. Sutherland, 1965. The Ultimate Display, In *Proc. Int. Federation for Information Processing,* p 506 - 508
[7] T. Hanna. What is somatics? In *Bone, Breath & Gesture: Practices of Embodiment.* D.H. Johnson, ed. North Atlantic Books, Berkeley, CA, 1995, 341–352.
[8] A. Williamson, 2014. Moving Spiritualities. In *Dance, Somatics and Spiritualities*, eds A. Williamson, G. Batson, S. Whatley, and R. Weber, Intellect, Bristol, UK
[9] L. Loke and T. Shiphorst. 2018. The Somatic Turn in HCI, *Interactions* 25, 5, 54-5863. https://doi.org/10.1145/3236675